\documentclass[10pt,preprint2]{aastex}
\usepackage{wasysym}

\shorttitle{Rings Around Transiting EGPs}
\shortauthors{Barnes \& Fortney}

\begin{document}

\title{Transit Detectability of Ring Systems Around Extrasolar Giant
Planets}
\author{Jason W. Barnes and Jonathan J. Fortney\altaffilmark{1}}
\affil{Department of Planetary Sciences}
\affil{University of Arizona}
\affil{Tucson, AZ 85721}
\email{jbarnes@c3po.barnesos.net, jfortney@arc.nasa.gov}
\altaffiltext{1}{Now at NASA Ames Research Center, Mail Stop 245-3, Moffett Field, CA  94035}

\newpage

\begin{abstract} 

We investigate whether rings around extrasolar planets could be detected from
those planets' transit lightcurves.  To this end we develop a basic theoretical
framework for calculating and interpreting the lightcurves of ringed planet
transits based on the existing framework used for stellar occultations, a
technique which has been effective for discovering and probing ring systems in
the solar system.  We find that the detectability of large, Saturn-like ring
systems is largest during ingress and egress, and that reasonable photometric
precisions of $\sim 1-3 \times 10^{-4}$ with 15-minute time resolution should
be sufficient to discover such ring systems.  For some ring particle sizes,
diffraction around individual particles leads to a detectable level of
forward-scattering that can be used to measure modal ring particle diameters. 
An initial census of large ring systems can be carried out using high-precision
follow-up observations of detected transits and by the upcoming NASA
\emph{Kepler} mission.  The distribution of ring systems as a function of
stellar age and as a function of planetary semimajor axis will provide
empirical evidence to help constrain how rings form and how long rings last.

\end{abstract}


\section{INTRODUCTION}

Present understanding of extrasolar planets resembles that of
solar system planets when Galileo first used the telescope for astronomy in
1610.  He discovered that the planets of our solar system variously display
disks, phases, moons, and ring systems \citep{Galileo.1610}.  Though the planets
he was studying had been known for thousands of years, before Galileo's
telescopic observations scientists knew only the character of their motions and
lacked a method for learning more.  Now that the orbital parameters for 120
planets orbiting other stars are available, we are looking for the breakthrough
techniques that will lead us to the next level of understanding for these new
extrasolar planets, just as Galileo's telescope did 400 years ago for the
planets known in ancient times.

We think that the most promising technique for characterizing extrasolar giant
planets in the near future is precision photometry.  

Photometric monitoring of stars has resolved one transiting planet detected by
radial velocity methods, HD209458b
\citep{2000ApJ...529L..45C,2000ApJ...529L..41H}, and three more detected by
ground-based transit surveys: OGLE-TR-56b \citep{OGLE-TR-56b.discovery.2003},
OGLE-TR-113b \citep{bouchy.very.hot.Jupiters.2004,Konacki.OGLE113.2004}, and
OGLE-TR-132b \citep{bouchy.very.hot.Jupiters.2004}.  Fitting the transit
lightcurve of HD209458b allowed \citet{2001ApJ...552..699B} to measure the size
of the planet's disk, which determined HD209458b to be a gas giant like
Jupiter.  Further observations of the wavelength dependance of HD209458b's
transit lightcurve revealed the presence of upper atmospheric sodium
\citep{2002ApJ...568..377C} and an extensive hydrogen envelope
\citep{Evaporating.HD209458b.2003}.

Extrasolar planets in orbits not in the plane of the sky (\emph{i.e.}, not
face-on) should show phases.  Though searches to date have not detected the
photometric signature expected for the phases of orbiting extrasolar planets,
these studies were able to place constraints on the reflective properties of
close-in giant planets
\citep{1999ApJ...522L.145C,2001astro-ph..12186,2002MNRAS.330..187C}.  

\citet{1999A&AS..134..553S} showed that large moons around transiting extrasolar
planets should reveal themselves both by additional stellar dimming while they
transit and by their effect on the transit timing of their parent planets. 
Using transit photometry from the \emph{Hubble Space Telescope},
\citet{2001ApJ...552..699B} employed this technique to search for moons orbiting
HD209458b, but found no evidence for any, consistent with theory for the orbital
evolution of such moons \citep{ExtrasolarMoons}.

The purpose of the present work is to predict what observational photometric 
transit signatures a ringed planet might make and to point out the possible
scientific value of a survey for such signatures.  It is not intended to be a
final or comprehensive theory for the study of extrasolar rings in transit.  It
is intended to serve as a guide for observers, so that they might have insight
into what the unusual deviations in their transit lightcurve residuals might
be.

In this paper, we first investigate the possible science that could be done
from a transit photometric ring survey.  Next we study the practicality of
detecting systems of rings around extrasolar giant planets using transit
photometry by looking into the effects of extinction and diffraction.  Lastly,
we simulate a transit of Saturn as a well-known example.

\section{MOTIVATION}\label{section:motivation}

\subsection{Rings in the Solar System}

Ring systems orbit each of the giant planets in our solar system.  The rings
are made up of individual particles orbiting prograde in their host planet's
equatorial plane.  However, each system of rings is unique, differing in
character, radial and azimuthal extent, optical depth, composition, albedo, and
particle size.  See Table \ref{table:ring.info} for a brief summary of what is
known about the rings of our solar system.  

Among solar system planets, by virtue of its rings' large effective
cross-sectional area (which we define to be the product of the rings' actual
cross-sectional area and one minus their transmittance) Saturn by far has the
greatest capability to block starlight during a transit.  Thus for the
remainder of this paper we will concentrate on ring systems with similar radial
extent and optical depth, which we refer to as Saturn-like ring systems.  The
ring systems of Uranus, Neptune, and Jupiter would reamin undetectable with
current or predicted future photometric capabilities, and thus our analysis is
only applicable to the aforementioned large, Saturn-like ring systems.

\begin{deluxetable}{l|cccc}
\tablecaption{Characteristics of Known Ring Systems\label{table:ring.info}}
\tablewidth{0pt}
\tablehead{
\colhead{Planet} &
\colhead{Ring} &
\colhead{Ring} & 
\colhead{Ring} &
\colhead{Ring} \\
\colhead{} &
\colhead{Particle Size} &
\colhead{Composition} & 
\colhead{System Age} &
\colhead{Origin}
}
\startdata
Saturn &	0.01-10 m  & water ice& young?       & unknown \\ \hline
Uranus & not yet measured & uncertain     & unknown & unknown \\ \hline
Neptune& not yet measured & uncertain     & unknown & unknown \\ \hline
Jupiter& mostly     & silicate & continuously & hypervelocity impacts  \\
       & $< 1\mu m$ &   dust   & replenished  & into adjacent moonlets \\
\enddata
\end{deluxetable}

\subsection{Rings in Extrasolar Systems}

A survey of ring systems orbiting extrasolar giant planets can potentially
address the two big picture questions about ring systems:  (1) how do rings
form? and (2) how long do rings last?  

If Saturn's ring system formed recently and ring systems decay quickly, then
extrasolar ring systems as spectacular as Saturn's should be rare.  If ring
systems around extrasolar giant planets are shown to be common, then ring
systems are either long-lived, easily and frequently formed, or both.

The typical lifetime of ring systems can be tested directly based on the
distribution of ring systems detected around stars of differing ages.  If ring
systems survive for many billions of years, then the frequency of ring systems
as a function of stellar age should be either flat (if rings are primordial) or
increasing (if they are formed by later disruptive events).  If rings are less
prevalent around planets orbiting older stars, then typical ring lifetimes can
be constrained by the observations.  

Uncertainties in stellar age measurements are notoriously large, though, and
those uncertainties will propogate into any subsequent estimate of ring
formation times and lifetimes using our method.  It seems reasonable that
10 - 20 ringed planets would be necessary to attempt this type of analysis -- a
number possibly in excess of the number of ringed planets \emph{Kepler} alone
will discover, depending on what fraction of planets have Saturn-like ring
systems.  On the other hand, Kepler might discover that a large fraction of
very young (< 1 Gyr) planets have ring systems, thus improving the statistics
and revealing a primordially created population of rings.  At present we lack
the basic understanding that would be required to create a model to predict
the statistics for ring distribution as a function of stellar age.

Photometry during stellar occultations revealed the Uranian and Neptunian ring
systems;  photometry of extrasolar giant planet transits can be used to search
those planets for rings.  Note, though, that an extrasolar ring census could
also be done using photometry of reflected light from directly detected planets
\citep{2003DPS....35.1804D,2004astro.ph..3330A} or by microlensing
\citep{Gaudi.microlensing.spots.2003}.

Planets with small semimajor axes are most likely to transit.  However, ring
systems around the closest-in planets (semimajor axis less than $\sim 0.1$ AU)
are expected to decay over timescales that are short compared with the age of
the solar system, due to such effects as Poynting-Robertson drag and viscous
drag from the planet's exosphere \citep{Gaudi.microlensing.spots.2003}. 
Although the currently ongoing transit searches are aimed at finding only these
short-period planets \citep[][]{2003astro.ph..1249H}, the \emph{Kepler}
\citep[see][and references therein]{Kepler.reflected.light.2003} space mission
will monitor a single region of the sky for up to 4 years and be capable of
detecting transiting planets with any period. 

Planets with periods greater than the length of a photometric survey cannot
have their periods determined, and thus cannot be followed up on.  As we show
in this paper, though, the \emph{Kepler} mission should attain sufficient
temporal resolution and photometric precision to detect Saturn-like ring
systems without additional observations.  The number of transiting planets at
10 AU that we expect to be detected is low.  Using the approximate detection
probability  $R_*/a_p \times 2/\pi \times T_{\mathrm{obs}}/P_{\mathrm{planet}}$
(with $R_*$ as the stellar radius and $a_p$ the planet's semimajor axis) for
planets with periods ($P_{\mathrm{planet}}$) longer than the observation time
($T_{\mathrm{obs}}$), if every star had a Saturn then \emph{Kepler} would find
$\sim5$ among the 100,000 stars it will monitor.  If all Saturn-like ring
systems are made of water ice, then only transiting planets at large semimajor
axes could posess rings and therefore the statistics for a ring survey of
extrasolar giant planets are not good.

However, every large ring system need not necessarily be made of water ice. 
Although the composition and temperature of Saturn's rings may yet prove
necessary for their formation and continued existence, we cannot at present
rule out silicate rings inward of the ice line, metallic rings around close-in
planets, or rings made of lower temperature ices at larger planetary semimajor
axes.  \emph{Kepler} will find and search 50 HD28185b-type giant planets at
1AU, if every star has one.  The presence or absence of these unfamiliar ring
systems will help constrain the processes involved in the formation and
evolution of ring systems as a whole.

\section{EXTINCTION}

\subsection{Methods}

Rings affect a planet's transit lightcurve in two ways:  extinction and scattering. 
Multiple scattering is negligible in most cases.  Extinction occurs as a
result of the interception of incoming starlight by ring particles that either
absorb, reflect, or diffract the light, removing it from the beam.  The amount of
light transmitted through any given portion of the ring is equal to $e^{-\tau/\beta}$
where $\tau$ is the normal optical depth and $\beta$ is equal to the sine of the
apparent tilt of the rings (zero being edge-on).  The factor $\beta$ compensates for
the increased extinction at low tilt angles.

No presently existing or near-future technology can resolve other main-sequence
stars, much less a planet in transit across a star's disk.  Instead, we
calculate the total stellar brightness that would be observed during the transit,
and determine whether the structure of dimming expected for transiting planets
with rings could be differentiated from the structure expected for planets
without rings.

To determine the amount of expected dimming, we compare the integrated stellar
surface brightness to the flux intercepted by a planet and its potential ring
system.  The stellar surface brightness $I$ as a function of projected distance
from the star's center, $\rho$, is expressed in terms of  
$\mu=\cos(\sin^{-1}(\rho/R_*))$ as in \citet{oblateness.2003}, where $c_1$ and
$c_2$ are coefficients that parameterize limb darkening:
\begin{equation}
\label{eq:limbdarkening}
\frac{I(\mu)}{I(1)} = 1 - c_1\frac{(1-\mu)(2-\mu)}{2}+c_2\frac{(1-\mu)\mu}{2}
~\mathrm{.}
\end{equation}
The constants $c_1$ and $c_2$ define the character of the limb darkening, with
$c_1$ measuring the overall magnitude and $c_2$ the second order shape.  This
nontraditional limb darkening parameterization allows meaningful representation of
stellar limb darkening from a single parameter, $c_1$, instead of two or more
\citep[see also ][ Section 3; our $c_1$ corresponds to their $u_1+u_2$]{2001ApJ...552..699B} and is therefore more appropriate for use in
fitting transit lightcurves.

We calculate the total stellar flux ($F_0$) by integrating $I(\mu)$ across the
disk of the star over the projected distance from the star's center, $\rho$:
\begin{equation}
\label{eq:Finf}
F_0 = \int_0^{R_*} \frac{2\pi\rho I(\rho)}{(R^\prime)^2}~d\rho  ~\mathrm{.}
\end{equation}
The distance from the star to the observer, $R^\prime$, drops out in Equation
\ref{eq:integralalgorithm} and so is not calculated explicitly.

To determine the fraction of starlight blocked by the planet or intercepted by ring
material, we integrate over both $\rho$ and the apparent position angle $\theta$:
\begin{equation}
\label{eq:Fblocked}
F_{blocked} = \int_0^{R_*} \int_0^{2\pi} 
\frac{I(\rho)~(1-T(\rho,\theta))}{(R^\prime)^2}\rho~d\theta~d\rho  ~\mathrm{,}
\end{equation}
where $T(\rho,\theta)$ is the fraction of light transmitted at the corresponding
$\rho$, $\theta$ location.  The relative flux detected by an observer at time
$t$ is then 
\begin{equation}
\label{eq:integralalgorithm}
F(t) = \frac{F_0 - F_{blocked}}{F_0}~\mathrm{.}
\end{equation}
This algorithm is similar to the one that we used in \citet{oblateness.2003}, but
with an explicit azimuthal integral to incorporate ring effects.

As we demonstrated in \citet{oblateness.2003}, the difference between the transit
lightcurve of a given planet and the lightcurve of \emph{that same planet} with
an additional feature (such as rings or oblateness) is not an appropriate measure
of the detectability of the feature.  Because the parameters of the transit such
as the planet radius ($R_p$), stellar radius ($R_*$), impact parameter  ($b=
\rho_{\mathrm{min}}/R_*$), and stellar limb darkening ($c_1$) are initially unknown and must
be determined using a fitting process, the presence of the ring manifests itself
as an astrophysical source of systematic error in the measurement of these
quantities.  These errors act to simulate the transit of the ringed planet, and
thus diminish the ring detectability.  We then define the detectability of a ring
around a transiting planet to be equal to the difference between the ringed
planet's transit lightcurve and the lightcurve of the planet-only model that best
fits the ringed planet's lightcurve.

\subsection{Results}

\begin{figure} 
\plotone{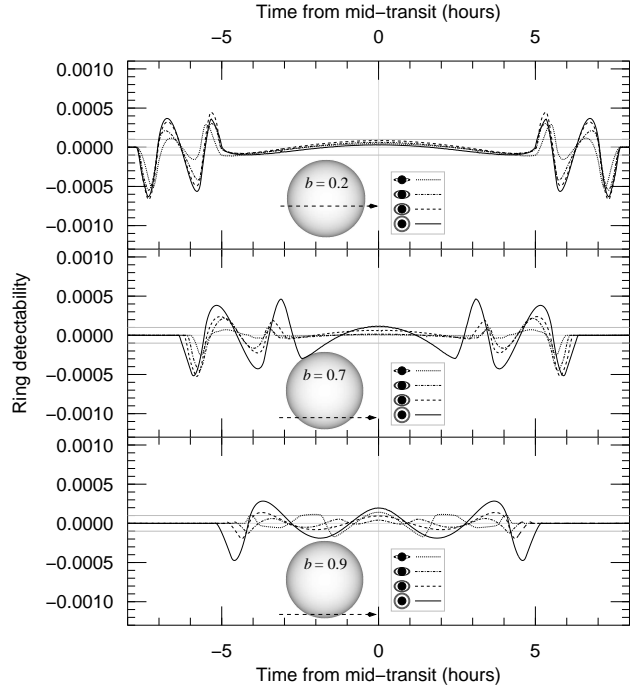}
\caption{Detectability of extinction through symmetric planetary rings in transit, 
defined as the difference between the transit lightcurve
of the given ringed planet and its best-fit spherical planet model.  This graph
shows the detectabilities for rings tilted directly toward the observer, Figure
\ref{figure:ring.asymm.detectability} shows the detectability for asymmetric
geometries.  Each
subgraph shows the detectability for planets of four different obliquities,
$10^\circ$ (dotted line), $30^\circ$ (dash-dot line), $45^\circ$ (dashed line), 
and $90^\circ$ (solid line;
face on) for
simulated transits with impact parameter 0.2 (upper), 0.7 (middle), and 0.9
(lower).  The signal is greater than the typical noise limit for both Kepler and
the HST HD209458b observations, $1\times10^{-4}$ (gray lines), but is very
localized in time to the regions surrounding ingress and egress.  Both high
photometric precision and high temporal resolution would be necessary to detect
the ring signal.
\label{figure:ring.symm.detectability}}
\end{figure}

\begin{figure}
\plotone{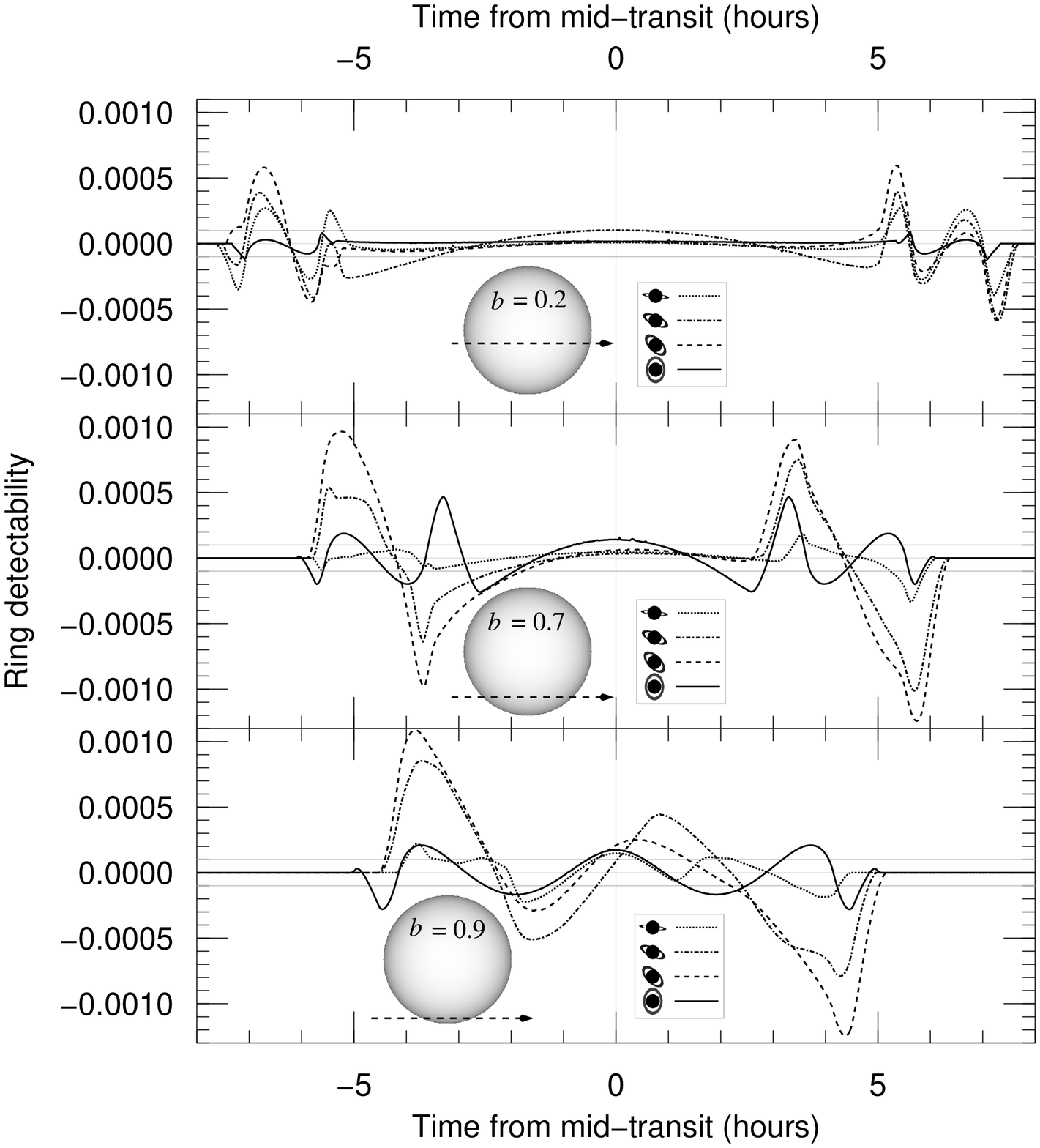}
\caption{Detectability of extinction through asymmetric planetary rings in
transit, defined as the difference between the transit lightcurve
of the given ringed planet and its best-fit spherical planet model.  This graph
shows the detectabilities for rings tilted in a direction such as to maximize
their transit detectability, so the axis angle $\phi$ is set to $\pi/4$.  Figure
\ref{figure:ring.symm.detectability} shows the detectability for symmetric
geometries.  Each
subgraph shows the detectability for planets of four different obliquities,
$10^\circ$ (dotted line), $30^\circ$ (dash-dot line), $45^\circ$ (dashed line), 
and $90^\circ$ (solid line;
face on) for
simulated transits with impact parameter 0.2 (upper), 0.7 (middle), and 0.9
(lower).  The signal is greater than the typical noise limit for both Kepler and
the HST HD209458b observations, $1\times10^-4$ (gray lines).  The asymmetric
signal is greater than that in the symmetric case, but only by a factor of
$\sim2$.
\label{figure:ring.asymm.detectability}}
\end{figure}

Figures \ref{figure:ring.symm.detectability} and
\ref{figure:ring.asymm.detectability} show the expected detectability (in extinction
only) of a planetary ring.  In these figures, we simulated transits of a spherical
$1R_{Jup}$ planet with a $\tau=1.0$ ring in its equatorial plane extending from
$1.5R_{Jup}$ to $2.0R_{Jup}$ in order to show the character that the detectability
of rings might have.  We address more realistic ring systems in Section
\ref{section:application}.  Although the particular lightcurves shown are for a
planet with a semimajor axis of 1.0 AU, \emph{the
detectability for planets at other distances from their stars is the same},
differing only in the timescale.  Thus, the results in this section apply equally
well for transiting planets that orbit their stars at $10\mathrm{AU}$ and for
close-in planets.  For a $1M_\odot$ star, to obtain the detectability curve of
a planet with semimajor axis $a_p$ multiply the values on the time axis by a factor
of $\sqrt{a_p/1.0\mathrm{AU}}$.  Transits of planets with differing radii and
ring structures will differ quantitatively from those plotted in Figures
\ref{figure:ring.symm.detectability} and \ref{figure:ring.asymm.detectability}.

For each case, the extinction of starlight through the rings leads to a deeper
transit, and therefore a transit lightcurve's best-fit ringless model planet is
larger than the real planet.  If an observer were to use the transit depth to
estimate the planet's parameters, he or she would overestimate the planet's
radius.  This could account for the anomalously high radius of HD209458b
\citep[e.g.,][]{Bodenheimer.planetradii.2003}.  However, we find this explanation
unlikely:  rings around HD209458b would be subject to strong orbital
perturbations due to their proximity to the star
\citep{Gaudi.microlensing.spots.2003}, and any ring system would be in the
planet's equatorial plane and therefore be seen edge-on during transit if, as
expected, the planet is in tidal equilibrium.

\begin{figure} 
\plotone{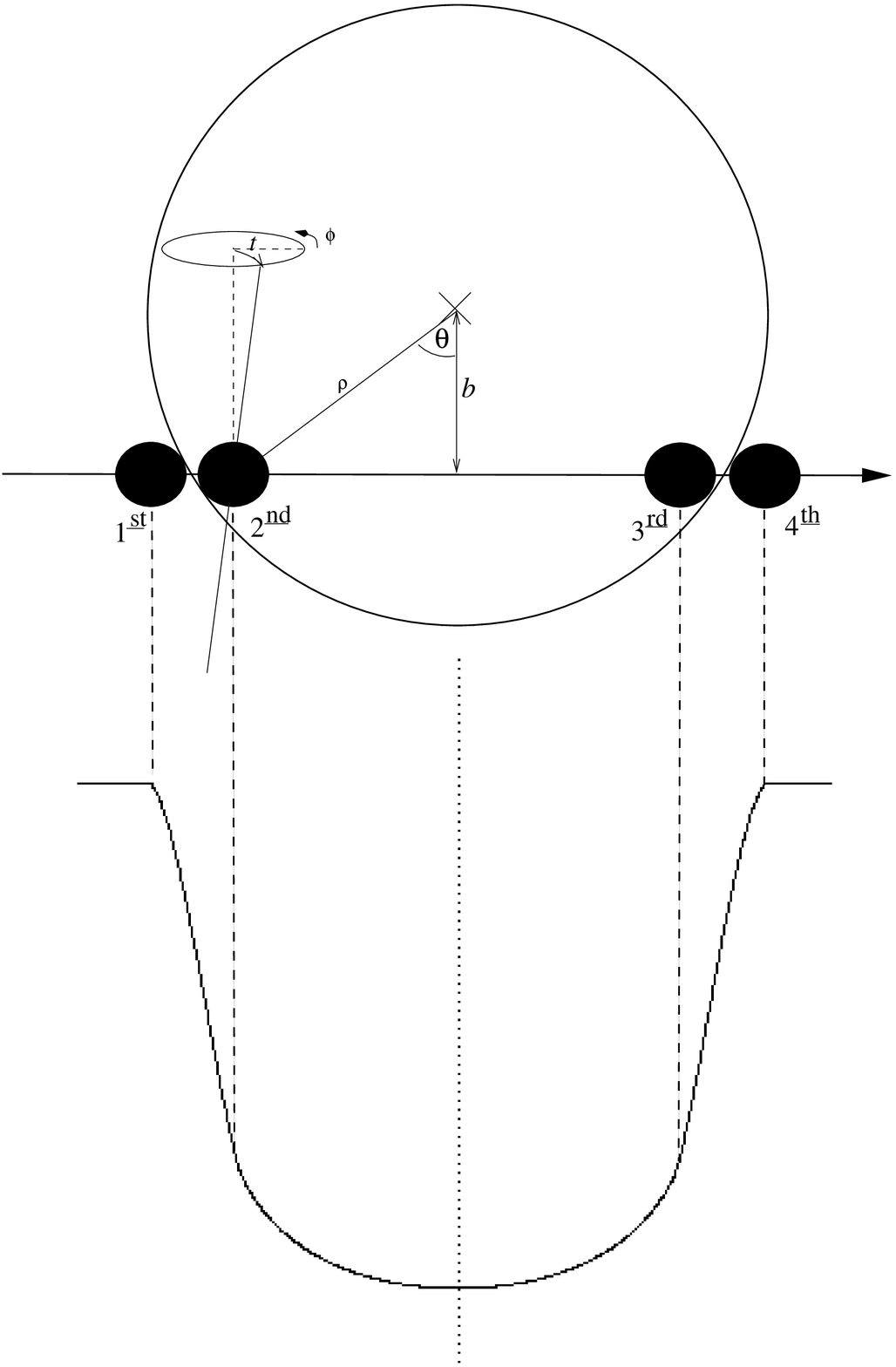}
\caption{Schematic of transit geometry and variable definitions.
\label{figure:schematic}}
\end{figure}

We introduce an angle $\phi$, which we call the axis angle, to represent the
azimuthal angle around the orbit normal corresponding to the direction that the
planet's rotational angular momentum vector points (see Figure
\ref{figure:schematic}).  We measure $\phi$ from the
direction that the planet is travelling in at mid-transit and positive is
defined to be toward the star.  The axis
angle thus corresponds to the planet's season at midtransit, with $\phi=0$ being the
northern fall equinox, $\phi=\pi/2$ being the northern summer solstice, $\phi=\pi$
being the northern spring equinox, and $\phi=3\pi/2$ being the northern winter
solstice.

We assume the rings to be two dimensional, which is quite a good approximation for
the rings of Saturn, Uranus, and Neptune:  Uranus' rings, for instance, are only
$1~\mathrm{m}$ thick \citep{2001Icar..151...78K}.  Hence, the rings we model do not at
all contribute to the lightcurve if they are edge-on during transit, and thus the
detectability for planets with obliquity ($q$) near $0^\circ$ or with axis angle
($\phi$) near 0 or $\pi$ is negligible.  Also, due to the geometry of the problem,
transits for a given axis angle and its opposite are identical.  Hence, the transits
of planets at solstice, such as those in Figure \ref{figure:ring.symm.detectability},
could have either $\phi=\pi/2$ or $\phi=-\pi/2=3\pi/2$.

In the symmetric case ($\phi=\pi/2$; Figure
\ref{figure:ring.symm.detectability}), for low impact parameter transits
($b=0.2$, upper subgraph) the lightcurve of the ringed planet starts earlier
than that of that transit's best-fit spherical planet as the rings encounter
the stellar limb first, leading to an initial downturn in the detectability
curve.  That curve turns around when the best-fit naked planet hits and the
ring's inner gap encounter the stellar limb at nearly the same time causing a
sharp change in derivative, whereby the best-fit lightcurve starts to overtake
the ringed planet in total light blocked.  As a large fraction of ringed planet
itself starts to occult the star, the two hypothetical transits become equal
about midway through their ingresses.  This trend continues past the midpoint
until the inner gap once again causes the simulated ringed planet transit to
brighten relative to its best-fit model, and by the end of ingress the edge of
the ring encounters the limb of the star at second contact.  The bottom of the
transit is nearly flat, as both the planets block the same fraction of
starlight.  This process repeats itself in reverse upon egress.

All things being equal, a non-ringed planet would have a much shorter ingress
and shallower transit depth than its ringed twin.  But all things are not
equal.  What happens instead is that the best-fit naked spherical planet model
mimics the simulated ringed planet's lightcurve by lengthening the model fit's
ingress and increasing the model's transit depth.  The resulting fits have a
higher impact parameter ($b$) and greater stellar radius ($R_*$), which both
act to increase the time that the model planet takes to cross the star's limb. 
At the same time, the model planet's radius ($R_p$) is higher than that of the
actual planet to account for the greater effective cross-sectional area that
the rings provide the ringed planet.  The stellar limb darkening coefficient
($c_1$) differs only slightly in this case to compensate for the different
transit impact parameters, and its inability to completely do so results in the
downward bow of the detectability curve through the transit bottom.

At this low impact parameter, the most important distinction between the ringed
planet transit and its best-fit model is the ringed planet's longer ingress, or
earlier first contact and later second contact.  Because a planet's rings extend
equally far from the planet's center regardless of obliquity, low obliquity rings
($q \sim 10^\circ$) are as detectable as high obliquity rings ($q >\sim 30^\circ$), but
remain distinctive by their smaller central clearing.  Though the ingress only
lasts about 45 minutes for a planet with HD209458b's semimajor axis, a planet at
1 AU would have a more leisurely (and more easily sampled) 3.75 hour ingress. 

The signature of a symmetric ringed planet is similar to, but distinct from, the
detectability signature of a highly oblate planet \citep{oblateness.2003}.  The
signature for a symmetric ringed planet returns to zero three times during
ingress, while the oblate planet detectability signature returns to zero just
twice.  The difference between the two is caused by the interior clear area
between the planet at the inner edge of the ring.  A planet with rings that extend
all the way down to the top of the atmosphere would be very difficult to
distinguish from a very highly oblate planet by observational means. 

At moderate impact parameters (\emph{i.e.}, $b=0.7$; Figure
\ref{figure:ring.symm.detectability}, middle subgraph), expected ring
detectabilities are similar in character and magnitude to those at low impact
parameter, but last longer due to longer ingress and egress times.  However,
symmetric, low obliquity ringed planets no longer have such extended ingress times
relative to a ringless planet, and thus have significantly lower detectabilities
than they do at lower impact parameters.

High impact parameter symmetric transits of ringed planets (\emph{i.e.}, $b=0.9$;
Figure \ref{figure:ring.symm.detectability}, bottom subgraph) deviate from the
patterns described above.  Planets with high obliquity have their ring systems
extend beyond the star's edge throughout the transit, and thus become grazing. 
Lower obliquity planets transiting at high impact parameter have their first
contact start no earlier than that of a ringless planet, and thus their
detectabilities are less than expected noise levels.

When a ringed planet is not tilted directly toward or away from the observer, the
lightcurve that results when the planet transits its star is asymmetric.  The
asymmetry is greatest for planets in mid-season, those with $\phi=\pi/4+n\pi/2$ where
$n$ is any integer, and the predicted detectability of rings around these planets is
displayed in Figure \ref{figure:ring.asymm.detectability}.  Differences from the
symmetric case come about due both to differing lengths of ingress and egress times
and to a different part of the planet-ring system making first contact with the stellar
limb.  These effects are most important when the planet's projected equatorial
plane (\emph{i.e.} the line containing the planet center and the apparent furthermost
edges of the rings) is parallel to the stellar limb during ingress of egress.  

At low impact parameter ($b=0.2$; Figure \ref{figure:ring.asymm.detectability}, top
subgraph), the planet's direction of motion is nearly perpendicular to the stellar
limb, and thus the differences between the symmetric and asymmetric lightcurves are
small, changing primarily the intensity of the positive and negative deviations but
not their character.  Small to moderate planetary obliquities ($10^\circ < q <
60^\circ$) therefore have transit lightcurves that strongly resemble the lightcurves from
the symmetric case.  The least detectable configuration at low impact parameter in
the mid-season case is for planets with obliquity $q=90^\circ$.  These Uranus-like
objects have symmetric transit lightcurves for all impact parameters, and low
detectabilities at $b=0.2$ because the stellar limb is covered by the rings at the
same time the rings' parent planet is covering the limb (a situation that the
best-fit nonringed planet simulates well).  

Differences between the time required for ingress relative to the time for egress
dominate the lightcurves for planets transiting near the critical impact parameter
($b=0.7$; Figure \ref{figure:ring.asymm.detectability}, middle subgraph).  The most
detectable cases are those where the projected equatorial plane is parallel to the
limb:  the $q=30^\circ$ and $q=45^\circ$ cases in Figure
\ref{figure:ring.asymm.detectability}.  For these ringed planets, first contact
occurs after, and second contact occurs before, the contacts for their best-fit
ringless counterparts in the case where the apparent planet motion is left to right
(see diagram at upper left of subfigures of Figure
\ref{figure:ring.asymm.detectability}.  Upon egress from transit, the rings encounter
the stellar limb before the best-fit spherical planet, and hang on to the limb after
the best-fit ringless planet ends its transit.  These differences in limb-crossing
time result in a large initially positive detectability (ringed minus best-fit
nonringed) as the best-fit planet starts its transit before the ringed planet, a
transition to negative values while both are near second contact, and then a return
to near-zero detectability for the transit bottom when both objects are entirely in
transit.  On egress, the resulting detectability is again initially positive as the
ringed planet lightcurve brightens when the rings exit transit before the limb of the
best-fit planet, and again becomes negative when the best-fit planet exits transit
before the lingering final edge of the rings of the ringed planet.  This character of
transit lightcurve is appropriate for planets with $-\pi/2 < \phi < \pi/2$.  For
planets with $\pi/2 < \phi < 3\pi/2$ the sequence would be time-reversed.  For low
obliquity ringed planets transiting at $b=0.7$, the asymmetric component of the
transit lightcurve is overshadowed by the symmetric component, and thus in character
resembles a cross between the two.  At maximum obliquity ($q=90^\circ$), the
detectability curve is symmetric but moderately large, with maxima resembling the
$\phi=0$ case in magnitude.

The detectabilities of asymmetric ringed planets when transiting at high impact
parameters ($b=0.9$; Figure \ref{figure:ring.asymm.detectability}, bottom subgraph)
are similar to those for $b=0.7$, except that the transit lightcurve sequences are
truncated due to the grazing nature of this particular transit.  Because there is no
second or third contact, there is no transit bottom and the detectabilities never
return to near zero as they do in the non-grazing cases.

The Discovery mission \emph{Kepler}, with typical photometric precision
$1\times10^{-4}$ and 15 minute sampling \citep{Kepler.reflected.light.2003}
should be most sensitive to planets with large ring systems and with periods
longer than $\sim1~\mathrm{yr}$, irrespective of scattering (see
Section \ref{section:diffraction}).  Higher cadence measurements with similar
precision, as acquired by \emph{HST} \citep{2001ApJ...552..699B} and as may be
possible in the near future from the ground \citep{Howell.OTCCD.preprint},
would be capable of detecting rings around closer-in planets, if such rings
exist (see Section \ref{section:motivation}).

\section{DIFFRACTION}\label{section:diffraction}

\subsection{Theory}

Calculation of scattering from a transiting extrasolar ring system resembles
the calculation of scattering from a ring in the solar system as it occults a
star.  In both problems, starlight passes through a thin slab of ring particles
and is altered on its way to the observer, who measures the star's brightness. 
However, while for solar system applications the distance between observer and
ring is small and the observer-star and ring-star distances are both large,
in the transit case the distance between the ring and the star is
small and the observer-star and observer-ring distances are large. 
\citet{French.Nicholson.2000} analyzed photometry from Saturn's occultation of
the star 28Sgr in 1989, and we base our theoretical models on theirs, with
appropriate modifications for the altered circumstances between occultations and
transits.

Particles between $1\mathrm{mm}$ and $10\mathrm{m}$ in size contribute most of
the opacity in the rings around Saturn, Uranus, and Neptune
\citep{French.Nicholson.2000, Cuzzi.1985, Ferrari.Brahic.1994}.  Particles like
these, much larger than the wavelength of light passing through them, scatter
light in the forward direction when incident light diffracts around individual
particles.  The relative scattered flux as a function of the scattering angle,
the phase function $P(\theta)$, is 
\citep[assuming spherical particles;][]{French.Nicholson.2000}
\begin{equation}
\label{eq:phasefunc}
P(\theta) ~ = ~ \frac{1}{\pi} \left( \frac{J_1(\frac{2\pi a}{\lambda}\sin\theta)}
                                                    {\sin\theta}
\right)^2 ~ \mathrm{,}
\end{equation}
where $a$ is the particle radius, $\lambda$ is the wavelength of light, and the
function $J_1()$ represents the first Bessel function.  Equation
\ref{eq:phasefunc} integrates to $1.0$ over the forward $2\pi$ steradians in
order to conserve flux, but is only accurate when $\theta \ll 2\pi$.  The shape
of this function resembles that of a stellar diffraction point spread function
from a telescope.  In effect, it represents the collective Poisson's spots of
the ring particles viewed from, say, 50 parsecs away.  

The total flux diffracted by the ring particles is equal to the flux that they
intercept.  This can be inferred from Babinet's Principle:  the diffraction of
a beam caused by a particular screen containing both transparent and opaque
areas appears the same as the diffraction of the screen's inverse, which has
the opaque and transparent areas switched \citep[see, e.g.,]
[the phase of the diffracted light is opposite in each case]{Hecht.Optics}. 
Hence the flux of light diffracted by a ring filled with particles of radius $a$
is the same as the flux of diffracted light coming from a black sheet peppered
with holes of radius $a$, which is equal to the light transmitted thorugh the
holes in the sheet.  Thus on top of absorption, diffraction provides an
additional, equal source of opacity for photons travelling straight through the
ring.  This effect can be confusing \citep{Cuzzi.1985}, so we follow
\citet{French.Nicholson.2000} in explicitly representing the optical depth
resulting from geometrically intercepted, absorbed light as $\tau_g$ and the
total optical depth, including scattering effects, as  $\tau ~ = ~ 2\tau_g$.  

The actual effective geometrical cross-sectional area of the ring particles
varies with optical depth due to shadowing effects.  We express the total
scattered flux that emerges from a point on the ring integrated over all angles
($F_{sc}^R$) as a function of the flux incident on that point ($F_i$)
\citep{French.Nicholson.2000}:
\begin{equation}
\label{eq:scatterfrac}
\frac{F_{sc}^R}{F_i} ~ = ~\frac{\tau_g}{\beta}e^{-2\tau_g/\beta} ~ \mathrm{.}
\end{equation}
$F_{sc}^R$ peaks at $\tau=1$ with a value $(2e)^{-1}$.  The variable $\beta$ is
equal to the sine of the apparent tilt of the rings (zero being edge-on), and 
compensates for the correspondingly higher optical depths for low tilt.  
The flux exitting the ring in the direction of the observer (designated
$\oplus$) at a specific angle ($\theta$) when diffracted from \emph{a point
source star}, $^\cdot F_{sc}^\oplus$, is the product of the total scattered flux and
the phase function, where $\theta$ corresponds to the angle between the
point source star-ring particle line and the line of sight \citep{French.Nicholson.2000}:
\begin{equation}
\label{eq:scatterflux}
^\cdot F_{sc}^\oplus(\theta) ~ = ~ F_{sc}^R ~ P(\theta)
\end{equation}

In the occultation case, the star behaves as a point source when viewed from
both Earth and the ring.  However, for transits the planet is in close enough
proximity to the star that the star must be treated as an extended source.  
Hence, the scattered flux from a point on the ring is not simply the product of
the incident flux on the ring and the phase function, as in the solar system
case, because $\theta$ is not constant.  Instead we integrate over the stellar
disk to derive the total scattered flux at Earth from across the extended star,
$^\circ F_{sc}^\oplus$.

We first calculate the illumination provided for each parcel of the star onto
the ring. To determine this incident flux, we treat the star as a limb-darkened
lambertian disk.  The flux incident on a point in the rings from a region of
that disk depends on the emergent stellar flux from the region as a function of
the projected distance from the star's center, $F_*(\rho)$ (which we get from
Equation \ref{eq:limbdarkening}); the area of the region, $E$; and the distance
between the region and the ring portion in question, which we approximate to
always be $a_p$, the planet's semimajor axis:
\begin{equation}
\label{eq:incidentflux}
F_{i}(R,\rho,\phi) ~ = ~ F_*(\rho) ~ \frac{E ~ \cos^2(\theta(R,\rho,\phi))}{\pi a_p^2} ~ 
\mathrm{.}
\end{equation}

One of the two $\cos\theta$ terms derives from our assumption that the star is
a flat disk, and the other comes from the need to measure the incoming flux in
a plane perpendicular to the observer's line of sight.  Though the second
$\cos\theta$ is real, the first term is an artifact of our model, and as such
is probably not representative of an actual planet around a spherical star. 
However, in the transit situation $\theta$ is small and the extra $\cos\theta$
does not inflict a significant adverse effect on the resulting calculation.

To obtain the total scattered flux from the extended star, we integrate the
scattered flux for a point source, Equation \ref{eq:scatterflux}, over the
projected distance from the star's center ($\rho$) and the azimuthal angle
($\phi$), using Equation \ref{eq:incidentflux} for the incident flux $F_i$. 
The total flux emergent from the ring, scattered from the extended star in the
direction of the observer ($^\circ F_{sc}^\oplus(R)$), comes from integrating
the star's flux over each infinitessimal area of the star $E=\rho~d\phi~d\rho$, times
the phase function for the particular angle $\theta$ between that area, the
scatterer, and the observer:
\begin{equation}
\label{eq:starfluxintegral}
^\circ F_{sc}^\oplus(R) ~ =
\int_0^{R_*}\hspace{-3mm}\int_0^{2\pi} ~ F_{sc}^R(\rho,\theta)~P(\theta(R,\rho,\phi))
\rho~d\phi~d\rho~
\mathrm{.}
\end{equation}
Due to circular symmetry around the center point of the limb darkened star, the
\emph{total} integrated incident flux is only a function of the projected
distance of the ring from the star's center, $R$, and not of the azimuthal
location of the rings around the star, $\phi$.  When incorporating scattered
light into the calculation of transit lightcurves, we integrate Equation
\ref{eq:starfluxintegral} numerically using an adaptive stepsize Runge-Kutta
algorithm from \citet{NumericalRecipes}.  

\subsection{Results} \label{section:results}

\begin{figure} 
\plotone{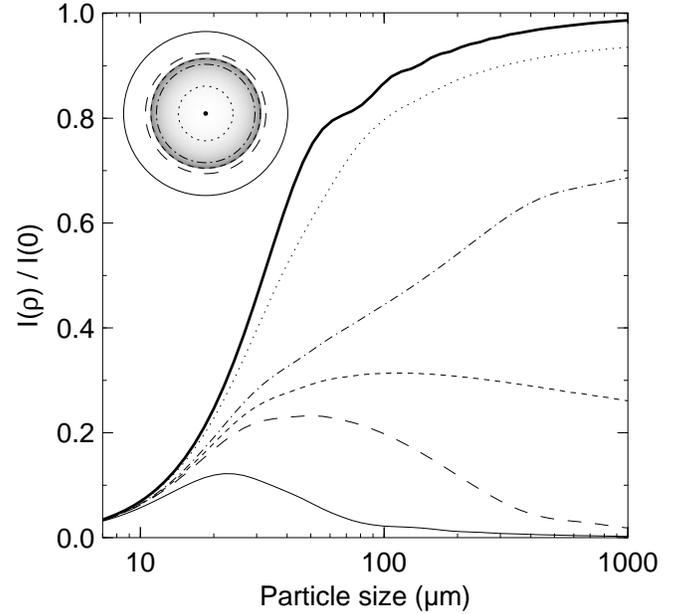}
\caption{This plot shows scattering effectiveness as a function of particle size at
$0.5 \mathrm{\mu m}$ for hypothetical parcels of ring particles located 1 AU from
the star but at varying projected distances from the star's center.  We used a
stellar diameter of $1.0 R_\odot$ and limb darkening coefficient $c_1$ equal to
0.64 (similar to the measured value for HD209458); the limb darkening in the inset
graphic is true.  We plot for the projected distances depicted in the graphic inset
in the upper left:  $\rho = 0.0 R_*$ (thick solid line),  $\rho = 0.5 R_*$ (dotted
line), $\rho = 0.9 R_*$ (dot-dashed line), $\rho = 1.0 R_*$ (short-dashed line), 
$\rho = 1.1 R_*$ (long-dashed line), and $\rho = 1.5 R_*$ (thin solid line).  We
have not included the effects of optical depth on the amount of light scattered --
to get the actual expected scattered flux multiply the values in this graph by the
$\tau$-factor, the right side of Equation \ref{eq:scatterfrac}.  At low particle
diameters, the scattering effectiveness approaches a constant and diminishing
value for all projected stellar distances.  For high particle diameters, the
effectiveness approaches the relative flux of the point on the star directly behind
the cloud of particles -- zero for points off the limb of the star, and following
the limb darkening while inside the limb.  For particle diameters near the critical
particle diameter $a_\mathrm{crit}$ (see Equation \ref{eq:acritical};
$a_\mathrm{crit}$ for the conditions used to generate this graph is $70\mathrm{\mu
m}$), the brightness for clouds of ring particles sitting just off the star's limb
is maximized.
\label{figure:partsizescattering}}
\end{figure}

\begin{figure} 
\plotone{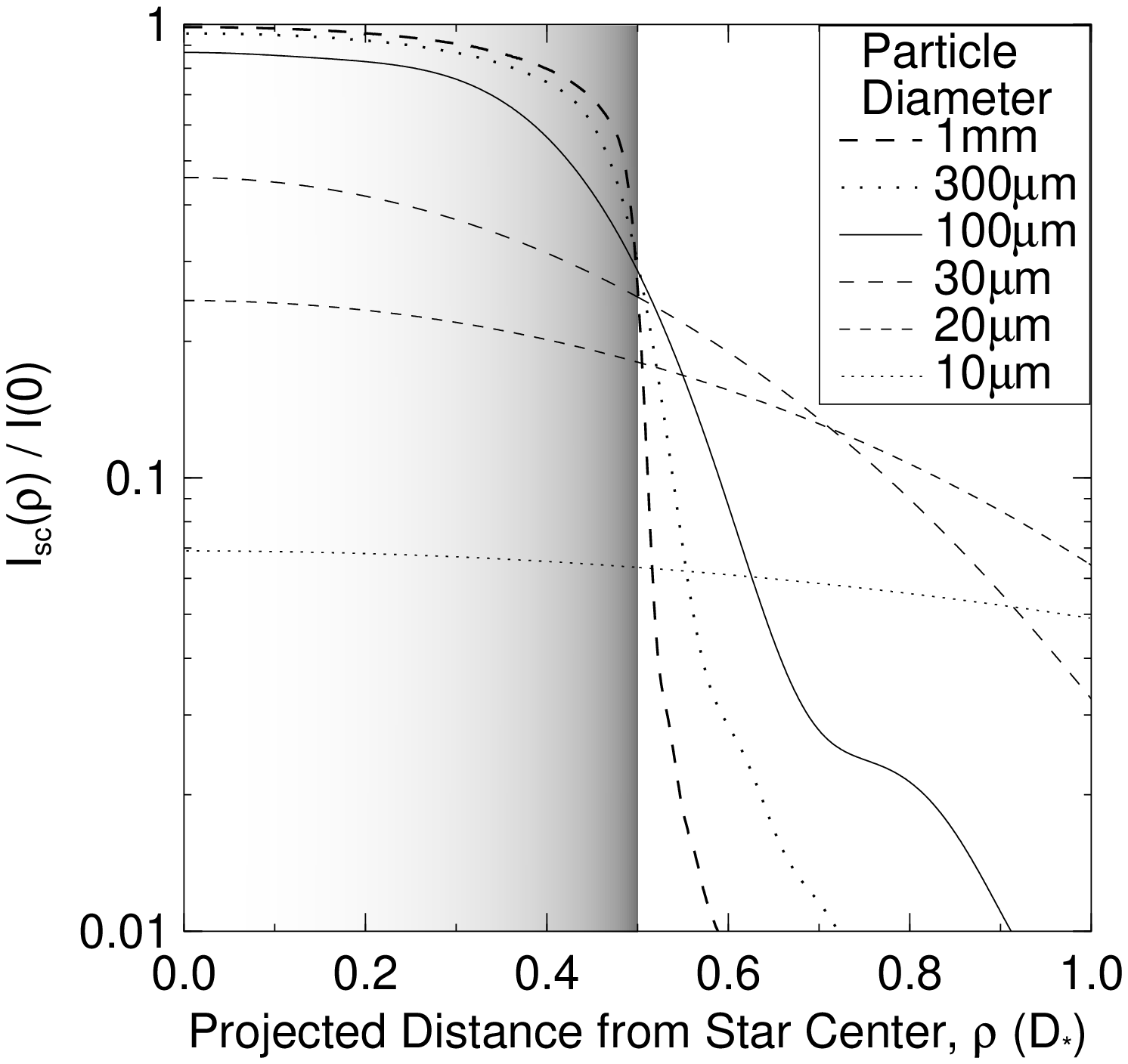}
\caption{Here we graph the scattering efficiency calculated in the same way as it
was for Figure \ref{figure:partsizescattering}, but do so in a different way, as a
function of projected distance from the star center and for different particle
diameters.  The shading in the background represents the actual limb darkened
stellar flux.
\label{figure:Rpositionscattering}}
\end{figure}

We show the relative contribution of diffracted light through a cloud of same-size
particles in Figures \ref{figure:partsizescattering} and
\ref{figure:Rpositionscattering}.  The amount of refracted light per unit projected
area relative to the flux per unit area at the star's apparent center is plotted as
a function of particle size for differing projected distance from the star center
(Figure \ref{figure:partsizescattering}), and as a function of projected distance
from the star center for differing particle sizes (Figure
\ref{figure:Rpositionscattering}).  For each we assume the planet's semimajor  axis
is 1.0 AU and that its parent star has a radius of $1~\mathrm{R_\odot}$ with limb
darkening coefficient $c_1$ equal to 0.64 ($(u_1+u_2)$ as measured by
\citet{2001ApJ...552..699B} for HD209458b) for the star.  The optical depth factor
from Equation \ref{eq:scatterfrac} is not included --  to obtain $^\circ
F_{\mathrm{sc}}^\oplus$ for any particular optical depth multiply the values on the
graph by the right side of Equation \ref{eq:scatterfrac} .

In the limit of very large particle sizes ($\sim 0.1$ mm or larger for the 1 AU case),
the angles by which light is diffracted around particles becomes very small.  These
very large particles diffract the light only from the point directly behind them on
the star toward the observer, and therefore in the large particle limit the maximum
diffracted flux per unit area at a given point is equal to the flux emitted by the
star at the point behind the ring (times the optical depth factor).  Hence to the
rightward edge of Figure \ref{figure:partsizescattering} the scattered fluxes
approach the flux levels at the equivalent point on the stellar disk.  In Figure
\ref{figure:Rpositionscattering}, the flux scattered by a sufficiently large
particle would trace the limb darkening of the star for $\rho < 0.5 D_*$ and be zero
for all values larger than $0.5 D_*$, similar to the curve of the 1 mm particles
plotted.  

\begin{figure} 
\plotone{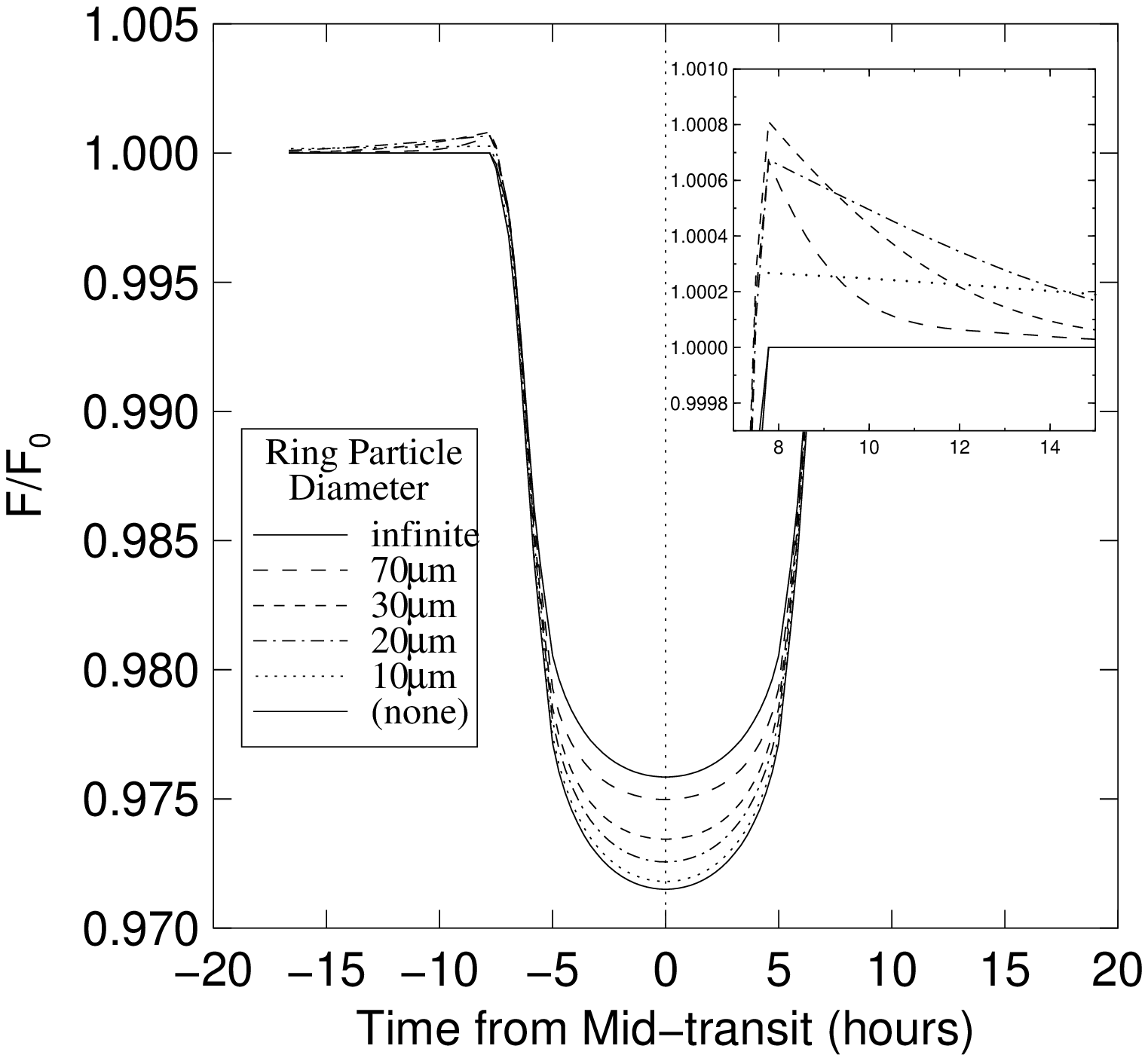}
\caption{Here we show the changes introduced into a transit lightcurve when
diffractive forward scattering is incorporated into the calculation.  This
figure shows transit lightcurves at $0.5\mathrm{\mu m}$ for a 1 $R_J$ planet with a 1.5 $R_J$ to 2.0
$R_J$ optical depth unity ring composed of particles of varying sizes.  The
planet orbits its 1 $R_\odot$, $c_1 = 0.64$ star at 1 AU, and transits with an
impact parameter of 0.2.  The bottom solid line represents what the lightcurve
calculated with extinction only -- no scattering.  The rest of the curves do
incorporate scattering along with extinction, and do so for differing particle diameters:  $10 \mu m$
(dotted line), $20 \mu m$ (dash-dotted line), $30 \mu m$ (short-dashed line), 
$70 \mu m$ (long-dashed line), and the theoretical (though meaningless in
practice) limit of infinite particle
size (top solid line).
\label{figure:scatter.lightcurve}}
\end{figure}

For a transiting ring made up of large particles, diffraction does not affect the
shape of the lightcurve.  However because light diffracted out of the incident
beam is redirected to the observer, diffraction would act to partially fill in the
lightcurve during the transit, reducing the total transit depth (Figure
\ref{figure:scatter.lightcurve}).  The reduced depth gives the impression that the
rings are dispersing less flux than they really are.  Thus, if the ring particles
are large, the \emph{measured} optical depth of the rings is equal to the
geometrical optical depth, $\tau_g = \tau/2$, because the light diffracted out of
the beam is entirely replaced (the same as if there were no diffraction at all).

For small particles (smaller than $\sim 10 \mathrm{\mu m}$ for the 1 AU case),
incoming light is diffracted nearly isotropically.  In this limit only a small
fraction of the incident flux is diffracted in any one direction, and the received
scattered flux is constant over the planet's orbit (except, of course, during the
secondary eclipse).  The diffracting behavior of such particles is represented by
the leftmost edge of Figure \ref{figure:partsizescattering} and resembles the
behavior of the $10 \mathrm{\mu m}$ particle in Figure
\ref{figure:Rpositionscattering}. 

Since for a ring made up of small particles the diffracted flux reaching the
observer is unvarying, the shape of the planet's transit lightcurve is the same
as it would be in the absence of scattering (Figure
\ref{figure:scatter.lightcurve}).  For this case the best-fit measurement of the
ring's optical depth would represent the total optical depth, $\tau = 2 \tau_g$: 
a negligible fraction of the diffracted light reenters the beam.  

Note that the large particle and small particle cases cannot be distinguished based
on their transit lightcurves alone.  If a ring were detected in transit, but no
distinctive signature of scattering was found (like are described below), it would be
impossible to determine whether the ring's component particles were large or small. 
Likewise the proper interpretation of the measured optical depth would also be
degenerate.

\begin{figure}  
\plotone{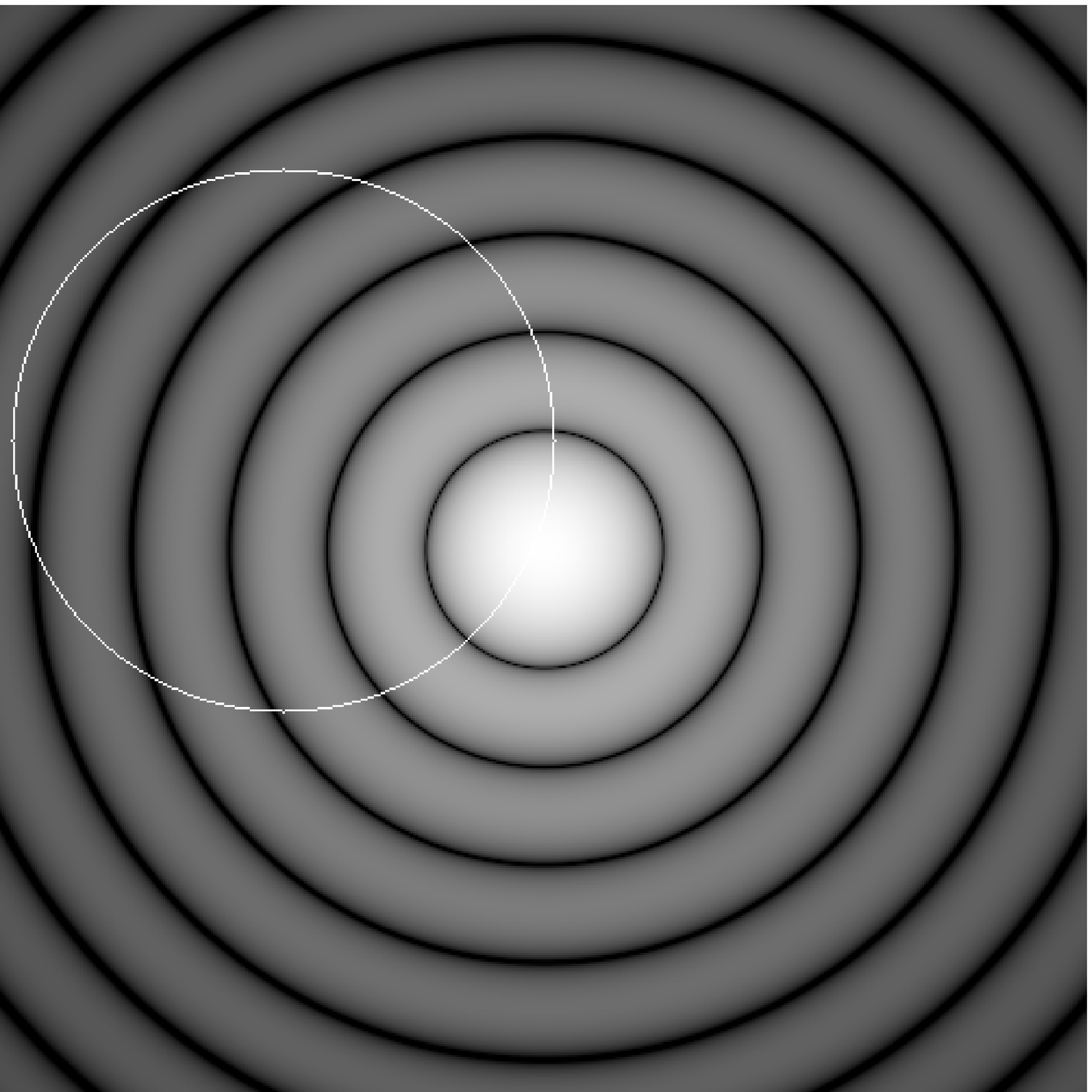} 
\caption{This image represents the diffraction footprint of a hypothetical 
cloud of $150
\mathrm{\mu m}$ particles (located at the center of the image)
1 AU from a 1 $\mathrm{R_\odot}$ star (the star's limb is depicted as a white
circle) as it would look at $0.5\mathrm{\mu m}$.  The intensity of each
pixel represents the logarithm of the relative contribution of light that
diffraction scatters toward the observer from each area.  The area directly
behind the particles contributes most of the diffracted light.  The effect of
diffraction on a ringed planet's transit lightcurve is most detectable when the
radius of the first dark fringe is $\sim 1 R_*$.  Since this process is
reversible, the same pattern would result if you fired a laser toward the
cloud of particles from Earth and then, hundreds of years later, viewed the resulting
diffraction pattern on a gigantic screen millions of km across
centered on the position of the star.  In practice, because of the nonuniform
radii and shape of the ring particles such a sharp pattern would never actually
occur. 
\label{figure:scatterimage}}
\end{figure}

For ring particles that lie in between these extremes of size, scattering affects the
lightcurve and can be used to infer the size of the component particles that make up
the ring.  In this moderate regime, the light diffracted toward the observer from a
parcel of ring particles comes from a circularly symmetric region behind the parcel
that grows in size as the particles' diameters shrink.  This footprint, shown in
Figure \ref{figure:scatterimage}, represents a weighted map of where photons
scattered toward the observer come from.  Most of the light comes from the area
within the first fringe.

The effect of scattering on a transit lightcurve is maximized for ring
particles near the size that has its first diffrataction fringe a projected
distance $R_*$ from the particles.  This critical particle size is large enough
to allow the footprint to scatter a lot of light toward the observer while the
ring is in front of the star's disk, but small enough to scatter significant
light toward the observer while the planet is off the limb of the star as
well.  We set $\theta$ to the first zero of the the phase function, $P(\theta)$
(Equation \ref{eq:phasefunc}), and solved to obtain the critical particle size,
$a_{\mathrm{crit}}$, which we define to be the ring particle radius near which
diffraction's effect on the lightcurve is maximized, using the small angle
approximation: \begin{equation} \label{eq:acritical} a_{\mathrm{crit}} = .61
\times \frac{a_p \lambda}{R_*} \end{equation}

Transit lightcurves for ringed planets change significantly as $a$ varies within an
order of magnitude of $a_{\mathrm{crit}}$.  However for particle diameters $a \gtrsim
10\times a_{\mathrm{crit}}$ and $a \lesssim a_{\mathrm{crit}}/10$, transit lightcurves
are indistinguishable from those for $a = 10\times a_{\mathrm{crit}}$ and for $a =
a_{\mathrm{crit}}/10$ respectively.  Thus large particles are those with $a \gg
a_{\mathrm{crit}}$, and small particles are those with $a \ll a_{\mathrm{crit}}$.  

Parcels of ring containing particles with diameters within a factor of $\sim 10$ of
$a_{\mathrm{crit}}$ scatter light to the observer both before and after they
encounter the stellar limb, as viewed from Earth.  This leads to an increased
photometric flux measured just before first contact, and just after fourth contact
during a transit event. The lightcurve signature (see Figure
\ref{figure:scatter.lightcurve}, inset) of this ramp is diagnostic of diffractive
scattering.  Moreover, the characterisic size of the extrasolar planet ring
particles can be determined from the shape and magnitude of the pre- and
post-transit scattering signature.

Only if the ring particle diameters are within a factor of 10 of $a_{\mathrm{crit}}$
can anything at all be discerned about the particle sizes -- otherwise, lightcurve
models using very large particles and those using very small particles will fit
equally well.

Similar lightcurve ramps to the ones shown in Figure
\ref{figure:scatter.lightcurve} can result from refraction through the outer layers
of the atmosphere of a transiting extrasolar giant planet. 
\citet{2001ApJ...560..413H} showed that refractive scattering will likely become
important only for planets with semimajor axes of tens of AU or greater.  In
particular, our calculations (using the codes developed for
\citet{2001ApJ...560..413H}) show that the refraction ramp for a typical transiting
EGP at 14 AU would be 100 times smaller than the ring diffraction ramps in Figure
\ref{figure:scatter.lightcurve}.  Therefore, a detected lightcurve ramp for a
transiting ringed planet can reasonably be ascribed to diffraction from ring
particles. 

Near $a_{\mathrm{crit}}$, rings' diffractive scattering affects the planet's
transit lightcurve in between first and fourth contacts as well.  As shown in
Figure \ref{figure:scatter.lightcurve}, the depth of the transit bottom
decreases as ring particle size increases because inside the transit larger
particles scatter light incident upon the ring into the direction of the
observer more efficiently than small particles do.  If $a\sim
a_{\mathrm{crit}}$, more of the diffraction footprint will overlap with the
star near mid-transit than at the beginning and end of transit.  In this case,
more diffracted light is detected at midtransit, and the transit lightcurve
bottom is flattened.  Fitting the simulated lightcurves plotted in Figure
\ref{figure:scatter.lightcurve} with spherical planets, we calculate that the
flattening is of order $\Delta c_1 \sim 0.1$ for a standard ring.

Though the lightcurve deviations described in this section are illustrative of
those we would expect to see owing to diffraction through planetary rings, they
are not a perfect calculation of what diffraction through a real ring system
will look like.  An actual ring system is likely to have complexity in its
radial structure that we have not simulated here.  Also, real ring particles
will have a nonuniform distribution of sizes, similar to the particles that
make up Saturn's rings \citep{French.Nicholson.2000}.  We are able to say,
however, that difraction around appropriately sized ring particles can lead to
a ramp in the transit lightcurve of a ringed extrasolar planet, that this ramp
can be detectable, and that such a ramp is greatly in excess of the expected
magnitude for similar ramps resulting from planetary atmospheric effects.

\section{APPLICATION}\label{section:application}

Saturn's ring system is distinctly more complex than the $\tau=1$, $1.5 - 2.0 R_{Jup}$
standard ring that we have used up to this point.  In particular,
\citet{2003DPS....35.1804D} showed that the intricate radial optical depth structure
of Saturn's rings significantly alters the directly-detected orbital lightcurve
relative to the lightcurve for radially uniform rings.  

To test whether the detectability of complex ring systems differs substantially from
that of our simple standard ring, we used a radial optical depth profile of Saturn's
rings obtained during the 28 Sgr occultation in 1989 \citep{28Sgr.occultation} binned
down into 39 subrings.  We set the rings' particle diameter to 1 cm, a typical value
from \citet{French.Nicholson.2000}.  We then simulated transits of Saturn at impact
parameter 0.2, calculated a best-fit spherical planet transit to match, and subtracted
the two to represent Saturn's detectability.  

\begin{figure}  
\plotone{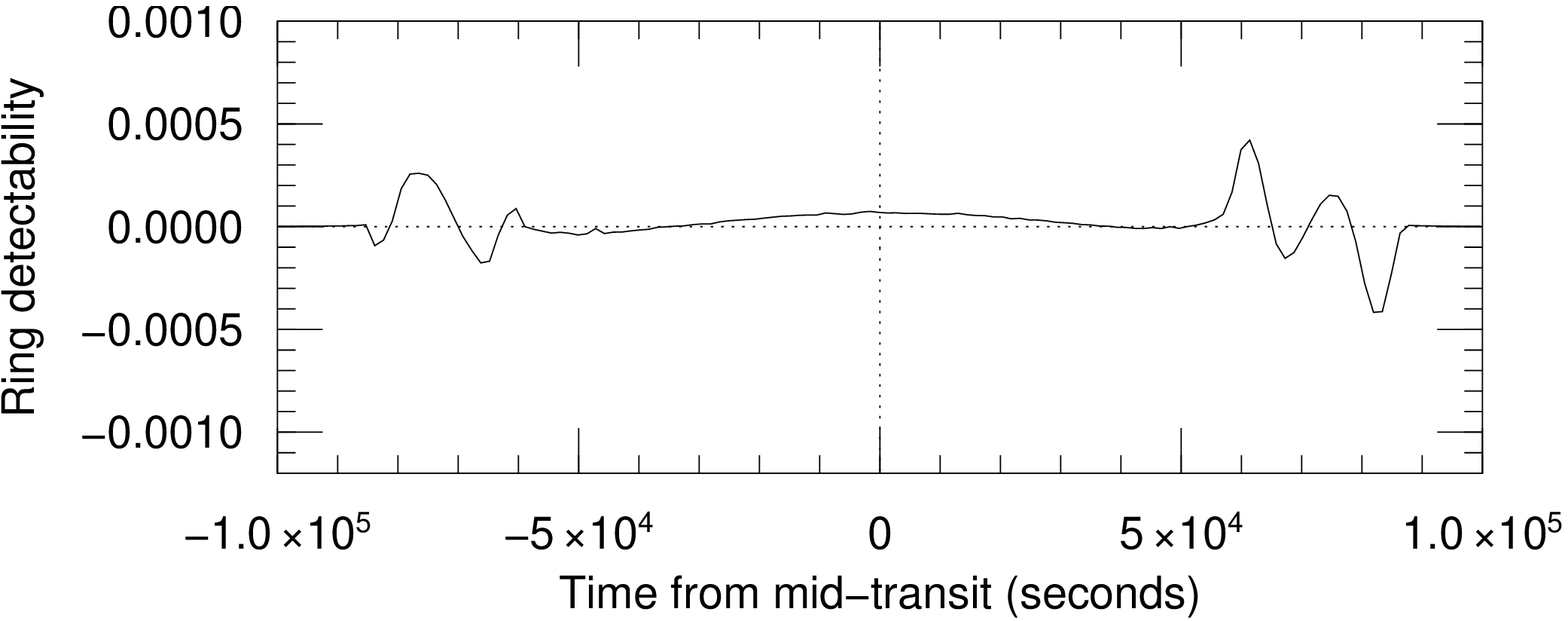}  
\caption{Predicted transit detectability of Saturn's rings, as might be viewed from 28
Sgr.  The magnitude is lower than that for the standard ring owing to the smaller
cross-sectional area of Saturn's rings, but the character of the signal is the same. 
No scattering is evident, so Saturn's ring particle size would not be discernable.
\label{figure:Saturn.detectability}}
\end{figure}

The detectability of the rings around Saturn, as shown in Figure
\ref{figure:Saturn.detectability}, is a bit less than the standard ring because the
total area covered is smaller.  However, the character of the ring signal is the same
as that of the standard ring.  Extrasolar astronomers, viewing a transit of Saturn
with a photometric precision of $10^{-4}$ from 28 Sgr, would detect Saturn's rings as
long as the axis angle $\phi$ wasn't such that they were viewed edge-on.

At 10 AU from the Sun, $a_{\mathrm{crit}}$ for Saturn is 0.66 mm (from Equation
\ref{eq:acritical}).  Therefore, as calculated Saturn's particle size falls in
the large regime and we would predict no scattering effects to be discernable. 
Figure \ref{figure:Saturn.detectability} does not show any significant pre- or
post-transit scattered light, in agreement with our analysis.  Without a positive
identification of scattering, the astronomers from 28 Sgr would not know whether
the ring particles were large or small, and therefore would not know whether
their measured ring optical depth corresponded to the geometric optical depth
($\tau_g$) or the total optical depth ($\tau$ -- see Section
\ref{section:results}).

\section{CONCLUSION}

The primary effect rings have on a planet's transit lightcurve is to increase the
transit depth.  Without knowledge of the rings' existence, the best-fit unadorned
spherical model planet will have an anomalously large radius and transit closer to
$b=0.7$ than the actual ringed planet.

The detection of large, Saturn-like ring systems requires photometric precision of a
few $\times 10^{-4}$, achievable with space based photometers and potentially from
future ground-based platforms.  The transit signature of rings is concentrated at
ingress and egress, and therefore ring detection requires high time-resolution
photometry.

Diffraction around individual ring particles manifests as forward scattering that can
be detected from transit lightcurves as excess flux just before and just after a
ringed planet's transit.  The shape and magnitude of the scattering signal can be used
to discern the ring particles' modal diameter.

The frequency and distribution of rings around planets at different semimajor axes, of
different ages, and with different insolations can empirically determine how rings
form and how long they last.

In the near-future, the experiments described in this paper can be tested on
close-in transiting planets found from the ground, using follow-up transit
photometry from space.  Later, the NASA Discovery mission \emph{Kepler} will be
capable of both detecting planets with larger semimajor axes, and surveying
these new worlds for rings.  Together, these observations hold the promise
to establish a rough outline of the distribution of large Saturn-like ring
systems within the next decade.

\acknowledgements

The authors wish to thank:  Robert H. Brown and William B. Hubbard for 
excellent advising; Renu Malhotra, Gwen Bart, and Wayne Barnes for manuscript
suggestions; and Peter Lanagan for allowing the generous use of his U of A
Press library.

\newpage


\newpage

\begin{thebibliography}{30}
\expandafter\ifx\csname natexlab\endcsname\relax\def\natexlab#1{#1}\fi

\bibitem[{{Arnold} \& {Schneider}(2004)}]{2004astro.ph..3330A}
{Arnold}, L. \& {Schneider}, J. 2004, A\&A, 420, 1153

\bibitem[{{Barnes} \& {Fortney}(2003)}]{oblateness.2003}
{Barnes}, J.~W. \& {Fortney}, J.~J. 2003, \apj, 588, 545

\bibitem[{{Barnes} \& {O'Brien}(2002)}]{ExtrasolarMoons}
{Barnes}, J.~W. \& {O'Brien}, D.~P. 2002, \apj, 575, 1087

\bibitem[{{Bodenheimer} {et~al.}(2003){Bodenheimer}, {Laughlin}, \&
  {Lin}}]{Bodenheimer.planetradii.2003}
{Bodenheimer}, P., {Laughlin}, G., \& {Lin}, D.~N.~C. 2003, \apj, 592, 555

\bibitem[{{Bouchy} {et~al.}(2004){Bouchy}, {Pont}, {Santos}, {Melo}, {Mayor},
  {Queloz}, \& {Udry}}]{bouchy.very.hot.Jupiters.2004}
{Bouchy}, F., {Pont}, F., {Santos}, N.~C., {Melo}, C., {Mayor}, M., {Queloz},
  D., \& {Udry}, S. 2004, \aap, 421, L13

\bibitem[{{Brown} {et~al.}(2001){Brown}, {Charbonneau}, {Gilliland}, {Noyes},
  \& {Burrows}}]{2001ApJ...552..699B}
{Brown}, T.~M., {Charbonneau}, D., {Gilliland}, R.~L., {Noyes}, R.~W., \&
  {Burrows}, A. 2001, \apj, 552, 699

\bibitem[{{Charbonneau} {et~al.}(2000){Charbonneau}, {Brown}, {Latham}, \&
  {Mayor}}]{2000ApJ...529L..45C}
{Charbonneau}, D., {Brown}, T.~M., {Latham}, D.~W., \& {Mayor}, M. 2000, \apjl,
  529, L45

\bibitem[{{Charbonneau} {et~al.}(2002){Charbonneau}, {Brown}, {Noyes}, \&
  {Gilliland}}]{2002ApJ...568..377C}
{Charbonneau}, D., {Brown}, T.~M., {Noyes}, R.~W., \& {Gilliland}, R.~L. 2002,
  \apj, 568, 377

\bibitem[{{Charbonneau} {et~al.}(1999){Charbonneau}, {Noyes}, {Korzennik},
  {Nisenson}, {Jha}, {Vogt}, \& {Kibrick}}]{1999ApJ...522L.145C}
{Charbonneau}, D., {Noyes}, R.~W., {Korzennik}, S.~G., {Nisenson}, P., {Jha},
  S., {Vogt}, S.~S., \& {Kibrick}, R.~I. 1999, \apjl, 522, L145

\bibitem[{{Collier Cameron} {et~al.}(2000){Collier Cameron}, {Horne}, {James},
  {Penny}, \& {Semel}}]{2001astro-ph..12186}
{Collier Cameron}, A., {Horne}, K., {James}, D., {Penny}, A., \& {Semel}, M.
  2000, ArXiv Astrophysics e-prints

\bibitem[{{Collier Cameron} {et~al.}(2002){Collier Cameron}, {Horne}, {Penny},
  \& {Leigh}}]{2002MNRAS.330..187C}
{Collier Cameron}, A., {Horne}, K., {Penny}, A., \& {Leigh}, C. 2002, \mnras,
  330, 187

\bibitem[{{Cuzzi}(1985)}]{Cuzzi.1985}
{Cuzzi}, J.~N. 1985, Icarus, 63, 312

\bibitem[{{Dyudina} {et~al.}(2003){Dyudina}, {Del Genio}, {Dones}, {Throop},
  {Porco}, \& {Seager}}]{2003DPS....35.1804D}
{Dyudina}, U.~A., {Del Genio}, A.~D., {Dones}, L., {Throop}, H.~B., {Porco},
  C.~C., \& {Seager}, S. 2003, AAS/Division for Planetary Sciences Meeting, 35,
  0

\bibitem[{{Ferrari} \& {Brahic}(1994)}]{Ferrari.Brahic.1994}
{Ferrari}, C. \& {Brahic}, A. 1994, Icarus, 111, 193

\bibitem[{{French} \& {Nicholson}(2003)}]{28Sgr.occultation}
{French}, R. \& {Nicholson}, P. 2003, NASA Planetary Data System,
  {USA\_NASA\_PDS\_EBROCC\_001}

\bibitem[{{French} \& {Nicholson}(2000)}]{French.Nicholson.2000}
{French}, R.~G. \& {Nicholson}, P.~D. 2000, Icarus, 145, 502

\bibitem[{{Galilei}(1989)}]{Galileo.1610}
{Galilei}, G. 1989, {Sidereus nuncius, or, The Sidereal messenger} (Chicago :
  University of Chicago Press, 1989.)

\bibitem[{{Gaudi} {et~al.}(2003){Gaudi}, {Chang}, \&
  {Han}}]{Gaudi.microlensing.spots.2003}
{Gaudi}, B.~S., {Chang}, H., \& {Han}, C. 2003, \apj, 586, 527

\bibitem[{{Hecht}(1998)}]{Hecht.Optics}
{Hecht}, E. 1998, {Optics} (Reading, MA: Addison-Wesley)

\bibitem[{{Henry} {et~al.}(2000){Henry}, {Marcy}, {Butler}, \&
  {Vogt}}]{2000ApJ...529L..41H}
{Henry}, G.~W., {Marcy}, G.~W., {Butler}, R.~P., \& {Vogt}, S.~S. 2000, \apjl,
  529, L41

\bibitem[{{Horne}(2003)}]{2003astro.ph..1249H}
{Horne}, K. 2003, ArXiv Astrophysics e-prints

\bibitem[{{Howell} {et~al.}(2003){Howell}, {Everett}, {Tonry}, {Pickles}, \&
  {Dain}}]{Howell.OTCCD.preprint}
{Howell}, S.~B., {Everett}, M.~E., {Tonry}, J.~L., {Pickles}, A., \& {Dain}, C.
  2003, \pasp, 0, 000

\bibitem[{{Hubbard} {et~al.}(2001){Hubbard}, {Fortney}, {Lunine}, {Burrows},
  {Sudarsky}, \& {Pinto}}]{2001ApJ...560..413H}
{Hubbard}, W.~B., {Fortney}, J.~J., {Lunine}, J.~I., {Burrows}, A., {Sudarsky},
  D., \& {Pinto}, P. 2001, \apj, 560, 413

\bibitem[{{Jenkins} \& {Doyle}(2003)}]{Kepler.reflected.light.2003}
{Jenkins}, J.~M. \& {Doyle}, L.~R. 2003, ArXiv Astrophysics e-prints

\bibitem[{{Karkoschka}(2001)}]{2001Icar..151...78K}
{Karkoschka}, E. 2001, Icarus, 151, 78

\bibitem[{{Konacki} {et~al.}(2003){Konacki}, {Torres}, {Jha}, \&
  {Sasselov}}]{OGLE-TR-56b.discovery.2003}
{Konacki}, M., {Torres}, G., {Jha}, S., \& {Sasselov}, D. 2003, \nat

\bibitem[{{Konacki} {et~al.}(2004){Konacki}, {Torres}, {Sasselov}, {Pietrzy{\'
  n}ski}, {Udalski}, {Jha}, {Ruiz}, {Gieren}, \&
  {Minniti}}]{Konacki.OGLE113.2004}
{Konacki}, M., {Torres}, G., {Sasselov}, D.~D., {Pietrzy{\' n}ski}, G.,
  {Udalski}, A., {Jha}, S., {Ruiz}, M.~T., {Gieren}, W., \& {Minniti}, D. 2004,
  \apjl, 609, L37

\bibitem[{{Press} {et~al.}(1992){Press}, {Teukolsky}, {Vetterling}, \&
  {Flannery}}]{NumericalRecipes}
{Press}, W.~H., {Teukolsky}, S.~A., {Vetterling}, W.~T., \& {Flannery}, B.~P.
  1992, {Numerical recipes in C. The art of scientific computing} (Cambridge:
  University Press)

\bibitem[{{Sartoretti} \& {Schneider}(1999)}]{1999A&AS..134..553S}
{Sartoretti}, P. \& {Schneider}, J. 1999, \aaps, 134, 553

\bibitem[{{Vidal-Madjar} {et~al.}(2003){Vidal-Madjar}, {Lecavelier des Etangs},
  {D{\' e}sert}, {Ballester}, {Ferlet}, {H{\' e}brard}, \&
  {Mayor}}]{Evaporating.HD209458b.2003}
{Vidal-Madjar}, A., {Lecavelier des Etangs}, A., {D{\' e}sert}, J.-M.,
  {Ballester}, G.~E., {Ferlet}, R., {H{\' e}brard}, G., \& {Mayor}, M. 2003,
  \nat, 422, 143

\end{thebibliography}
\end{document}